\RequirePackage{fix-cm}
\documentclass[twocolumn]{svjour3}          
\smartqed  
\usepackage{graphicx}
\usepackage{amsfonts}
\usepackage{amsmath}
\usepackage{bbm}
\usepackage{xcolor}
\usepackage{subfigure}  
\usepackage{enumerate, soul}

\DeclareFontFamily{OT1}{pzc}{}
\DeclareFontShape{OT1}{pzc}{m}{it}{<-> s * [1.10] pzcmi7t}{}
\DeclareMathAlphabet{\mathpzc}{OT1}{pzc}{m}{it}
\journalname{Nonlinear Dynamics}

\begin{document}

\title{Active control of liquid film flows: beyond reduced-order models
}

\author{Radu Cimpeanu         \and
        Susana N. Gomes        \and
        Demetrios T. Papageorgiou
}


\institute{R. Cimpeanu \at
              Mathematics Institute, University of Warwick, Coventry, CV4 7AL, UK \\
              Department of Mathematics, Imperial College London, London, SW7 2AZ, UK \\
              Mathematical Institute, University of Oxford, Oxford, OX2 6GG, UK \\
              Tel.: +44-24-76574809\\
              \email{Radu.Cimpeanu@warwick.ac.uk}           
           \and
           S.N. Gomes \at
              Mathematics Institute, University of Warwick, Coventry, CV4 7AL, UK \\
              \email{Susana.Gomes@warwick.ac.uk}
           \and
           D.T. Papageorgiou \at
              Department of Mathematics, Imperial College London, London, SW7 2AZ, UK \\
              \email{d.papageorgiou@imperial.ac.uk}
}

\date{Received: 24 September 2020 / Accepted: 3 February 2021}

\maketitle

\begin{abstract}

The ability to robustly and efficiently control the dynamics of nonlinear systems lies at the heart of many 
current technological challenges, ranging
from drug delivery systems to ensuring flight safety. Most such scenarios are too complex to tackle directly and reduced-order modelling 
is used in order to create viable representations of the target systems. The simplified setting allows for the development of rigorous 
control theoretical approaches, but the propagation of their effects back up the hierarchy and into real-world systems remains a significant challenge. 
Using the canonical setup of a liquid film falling down an inclined plane under the action of active feedback controls in the form of blowing and suction, 
we develop a multi-level modelling framework containing both analytical models and direct numerical simulations acting as an in silico experimental platform. 
Constructing strategies at the inexpensive lower levels in the hierarchy, we find that offline control transfer is not viable, however analytically-informed 
feedback strategies show excellent potential, even far beyond the anticipated range of applicability of the models. The detailed effects of the controls in 
terms of stability and treatment of nonlinearity are examined in detail in order to gain understanding of the information transfer inside the flows, which can 
aid transition towards other control-rich frameworks and applications. 

\keywords{Flow dynamics \and Reduced-order modelling \and Asymptotic analysis \and Stability \and Direct numerical simulation \and Control theory for distributed parameter systems \and Point-actuated feedback control \and Hierarchical modelling}
\end{abstract}

\section{Introduction}
\label{intro}

The ability to manipulate physical systems has applications in all areas of engineering, medicine, and many other fields. 
Examples range from path tracking and planning for self-driving cars~\cite{shum2015direction,zhao2012design}, 
in-flight measurement and manoeuvre adjustments in unmanned aerial vehicles (e.g. drones)~\cite{choi2018collisionless,oh2015survey},
management of crowds during mass gatherings or evacuations~\cite{bode2014human,johansson2012crowd}, 
to controlled release delivery systems for targeted drug delivery~\cite{tiwari2012drug}, to name but a few. 
However, before being able to control real-life systems, one must be able to model them accurately. 
There have been a number of recent investigative efforts on data-driven modelling and control of real-life systems, such as the derivation of 
coarse-grained partial differential equations (PDEs) from macroscopic observations~\cite{lee2020coarse}, the use of machine learning to predict behaviour and learn models in fluid mechanics~\cite{brunton2020machine}, or the design of controls based on models with uncertainty~\cite{kramer2017feedback}, to mention a few.
However, these settings do not allow for the rigorous theoretical development of control strategies and for these reasons 
we chose the reduced-order model approach that follows.
Real-life phenomena usually lead to extremely intricate models which are difficult to tackle analytically and 
are challenging computationally; 
this motivates the development 
of physically relevant reduced-order modelling approaches resulting in systems
that are amenable to mathematical and computational analysis. 
Typical examples include fluid dynamics~\cite{lassila2014model} and image restoration~\cite{angwin1989image}, see \cite{benner2015survey} for a recent review. 
The resulting models are usually accurate in certain regions of parameter space and are used as surrogates of the original system, thus
enabling detailed exploration of underlying phenomena without recourse to physical experiments and direct numerical simulations
({\it in silico} or virtual experiments).
Most importantly, such reduced-order models can be used as the starting point in controlling the dynamics of the more
general physical model, and this multi-way synergy between hierarchical models constitutes our present methodology.

This study has two main goals: (i) Establish the advantage of using control methodologies designed for reduced-order models in an 
{\it in silico} experiment; (ii) Use this framework as a test-bed for informing the control of large scale systems.
Our approach of using controls based on models lower in a hierarchy and moving up to the full physical problem, circumvents
challenges present in the latter class including lack of rigorous analytical results and reliance on expensive trial and error numerical experiments.
Proceeding directly with the full model could predict qualitative features in certain situations, but to our knowledge this is
the first attempt at propagating quantitative information between vastly different complexity levels.

The physical model we select to study is that of a
thin film of water flowing down an inclined plane. This has been studied extensively both theoretically and experimentally
and is ideal for our goals since accurate and efficient direct numerical simulations (DNS) of the Navier--Stokes equations
can be utilized as {\it in silico} experiments.
Furthermore, falling film flows have benefited from
a strong body of analytical efforts for reduced-order modelling
over the last 50 years (see, for example, \cite{benney1966long,sivashinsky1980irregular}; we also point out that the simplest model, 
the Kuramoto-Sivashinsky equation is also widely used as a model for flame front propagation, reaction-diffusion systems and 
other relevant scenarios \cite{kuramoto1975formation,sivashinsky1983instabilities,sivashinsky1977nonlinear}), 
as well as very recent comparisons between 
reduced-order models, full computations and experiments in relevant regimes \cite{denner2018solitary}.

Invoking the multiscale nature of the problem (the film is thin compared to its wavelength) leads to reduced-order 
long-wave or weakly nonlinear models \cite{kalliadasis2011falling}. These have been generalised to incorporate various effects 
that can act as active control mechanisms, 
such as electric fields~\cite{anderson2017electric,tomlin2020electric,tseluiko2006wave}, wall heating~\cite{blyth2012flow,thompson2019robust}
and same-fluid blowing and suction~\cite{thompson2016falling}. 
Alternative and/or supplementary passive control
can be achieved, for example, by including substrate topography~\cite{gaskell2004gravity,Sellier,TBP2013} 
or the inclusion of surfactants~\cite{blyth2004effect,Bontozoglou2018,Georgantaki2016,KB2013,KB2014}.
In this paper, we focus on same-fluid blowing and suction, which has recently been used to develop efficient feedback controls 
based on interface observations for a hierarchy of models of
increasing complexity~\cite{armaou2000feedback,gomes2017stabilizing,thompson2016stabilising,tomlin2019point}. 

Our results show that controls learned from reduced-order models based on observations of their interface are not directly applicable to the virtual experiment if they are imposed pre-experiment in the real-world system.
Nevertheless we can still obtain crucial information from the reduced-order models to be used in the full problem. 
The actuation rules (which are based on properties 
of the reduced-order models, often their linear stability properties) remain the same, only now the controls are predicated 
on observations of the interfacial shape in the 
virtual experiment. Furthermore, the predictions on the strength and number of controls needed to 
stabilise any desired solution, as well as their location 
and shift-displacement from observations is correct as demonstrated by extensive numerical simulations. 
The virtual experiments also allow us to explore the control methodology 
beyond the parameter regimes where the reduced-order models are valid, hence opening a new avenue for control development
learned from faster and parametrically rich computations. We emphasise that
our results and methodology are not restricted to the particular physical model considered here,
and are expected to be applicable to other similar systems, such as self-driving cars, smart robots, or active matter.

\section{The control methodology}
\label{s:methodology}

In this section, we outline the theory behind feedback control, which is the basis for the design of the controls developed in~\cite{gomes2017stabilizing,thompson2016stabilising}. We then present a hierarchy of models for the
dynamics of the interface of a thin film of water flowing down a plane in two dimensions (2D), which is the physical system of interest -- this hierarchy is obtained using asymptotic expansions that lead to reduced-order models. 

\subsection{An ODE example}
\label{ss:ODE}
The theory of linear feedback control was first established for (systems of) ordinary differential equations (ODEs), see~\cite{zabczyk1992mathematical}. To introduce this methodology we consider the simplest example of controlling a scalar ODE
\begin{equation}\label{e:toy-ODE}
\dot{y} = \lambda y, \qquad y(0) = y_0,
\end{equation}
where the dot represents derivative with respect to time. It is well known that the solution of \eqref{e:toy-ODE} is $y(t) = y_0 e^{\lambda t}$ and that $y(t)\rightarrow 0$ if $\lambda < 0$ and $y(t)\rightarrow\infty$ if $\lambda$ is positive. 

The main goal of linear control theory is to introduce a control to equation~\eqref{e:toy-ODE}, i.e.
\begin{equation}
\dot{y} = \lambda y + f, \qquad y(0) = y_0,
\end{equation}
and choose $f$ in such a way that the solution is \emph{stabilised}, \emph{i.e.}, driven to $y = 0$ even for $\lambda>0$. The simplest example of a \emph{proportional feedback control} would be to choose $f(t) = -\alpha y(t)$ for some positive constant $\alpha$. It is clear that if $\alpha$ is such that $\lambda-\alpha < 0$, then $y(t)\rightarrow 0$ as $t\rightarrow\infty$ and we say that the control $f(t)$ stabilises the solution to the ODE.
The term feedback is used because the control uses information on the current state of the system; the control is called \emph{proportional} since it is proportional to the current solution. 
A similar idea can be used in systems of ODEs
\begin{equation}\label{e:toy-system}
\dot{\mathbf{y}} = A \mathbf{y}  + \mathbf{f}, \qquad \mathbf{y}(0) = \mathbf{y_0},
\end{equation}
where now $\mathbf{y}, \, \mathbf{y_0}, \, \mathbf{f}\in\mathbb{R}^d$ and $A$ is a $d\times d$ matrix. Similarly to the one dimensional case, when $\mathbf{f}=\mathbf{0}$, if \emph{the eigenvalues} of $A$ all have negative real part, the solution is \emph{asymptotically stable}, i.e., $\mathbf{y}(t)\rightarrow \mathbf{0}$ as $t\rightarrow\infty$. The analogue of the previous control here is to use $\mathbf{f}(t) = -\alpha\mathbf{y}(t) = -\alpha I \mathbf{y}(t)$, where $I$ is the $d\times d$ identity matrix. A simple calculation can be used to find the smallest $\alpha$ necessary to stabilise the system,
namely, choose $\alpha$ such that the eigenvalues of $A-\alpha I$ all have negative real parts. 

It is naturally more efficient to choose the controls using more information about the model. This can be done by designing the control as follows
\begin{equation}
\label{e:feedbackeq}
\dot{\mathbf{y}} = A\mathbf{y}+B\mathbf{f}, \qquad \mathbf{y}(0) = \mathbf{y}_0.
\end{equation}
Here, $B$ is a $d\times M$ matrix that encodes some information about how one applies the controls (e.g., if the controls are localised in some part of the domain), and the controls are now $\mathbf{f}\in\mathbb{R}^M$. Note that we can have $M = d$ and $B = I$, which is the case outlined above. It can be shown that under certain assumptions on the matrices $A$ and $B$ (namely, the Kalman rank condition~\cite{zabczyk1992mathematical}), one can find a matrix $K$ such that the eigenvalues of $A+BK$ all have negative real part, and therefore the controls $\mathbf{f} = K\mathbf{y}$ stabilise the system. The matrix $K$ can be computed using, e.g., a pole placement algorithm~\cite{kautsky1985robust} or by solving a linear-quadratic regulator problem~\cite{zabczyk1992mathematical}.

In more realistic physical systems, the dynamics of the solution are modelled by a nonlinear system of ODEs,
\begin{equation}
\label{e:ode-nonlinear}
\dot{\mathbf{y}} = F(y) + B\mathbf{f}, \qquad \mathbf{y}(0) = \mathbf{y}_0,
\end{equation}
where $F$ is some nonlinear function of $y$. In addition, physical continuum models  provide (linear or nonlinear) \emph{partial differential equations} (PDEs), e.g., a reaction-diffusion equation for the evolution of a population, tumour growth or other biological and chemical applications. 
Such PDEs take the general form
\begin{equation}
\label{e:feedback_PDE}
u_t = \mathcal{L} u + \mathcal{N}(u) + f,
\end{equation}
along with appropriate initial and boundary conditions. The subscript $t$ denotes time derivative, and $\mathcal{L}, \, \mathcal{N}$ are linear and nonlinear spatial differential operators, respectively.
By projecting this equation to an appropriate basis (e.g., taking Fourier transforms), one can write the PDE as an infinite-dimensional 
system of ODEs such as~\eqref{e:ode-nonlinear}. Passing to this limit is not straightforward, even for linear PDEs~\cite{zabczyk1992mathematical}, however, in certain cases this is possible, even for nonlinear PDEs, as was shown for the  Kuramoto--Sivashinsky equation in~\cite{armaou2000feedback,gomes2017stabilizing}. In this case, in order to prove that the controls stabilise the full nonlinear system, one needs to use nonlinear stability analysis techniques, such as finding a Lyapunov function~\cite{kokotovic1999singular}.

\subsection{Reduced-order models for thin liquid films}
\label{ss:ROM}

\begin{figure}
\centering
\includegraphics[scale=0.32]{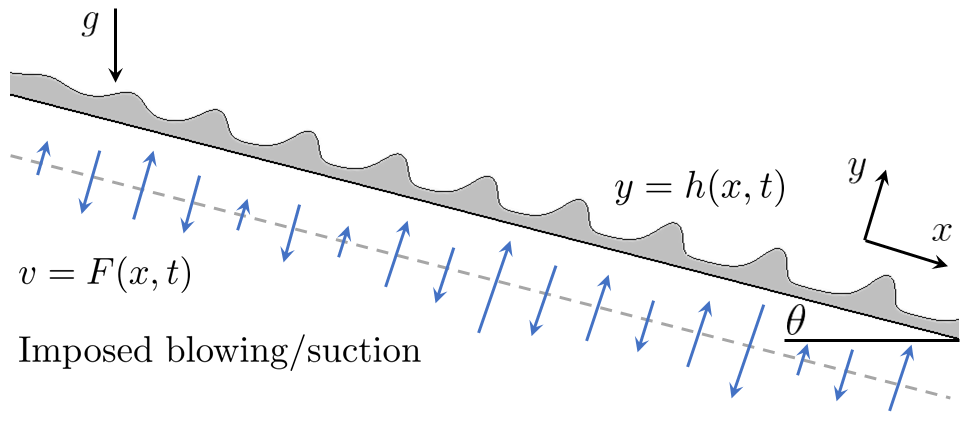}
\caption{Schematic of a falling liquid film down an inclined plane allowing for blowing and suction controls. The dynamics of the interface 
at $y=h(x,t)$ is controlled not only by fluid parameters, but also by the geometry inclination angle $\theta$ and the imposed control values 
$v=F(x,t)$ at $y=0$ in a suitably aligned coordinate system.}
\label{fig1}
\end{figure}

This section is devoted to a brief description of the physical problem, its mathematical modelling and the asymptotic work that
leads to the hierarchy of reduced-order models we are focussing on.
Consider a thin film of water flowing down a 2D plane inclined at an angle $\theta$ to the horizontal, 
see Figure~\ref{fig1}. 
Throughout this discussion, we will be using the $(\cdot)^*$ notation to distinguish dimensional quantities from their undecorated 
dimensionless counterparts. 
The film thickness is denoted by $h^*(x,t)$ and our main goal is to drive the system to its undisturbed flat interface, 
$h^*(x,t) = h^*_0$. We will apply proportional feedback controls which are actuated by means of blowing or suction of water via slots in the wall, 
and the controls will be designed based on readings of the interfacial height.

The full mathematical model for the fluid motion is given by the Navier--Stokes equations 
for both liquid film and gas above it with a suitable nonlinear coupling at the interface.
Due to the passive nature of the gas in our applications (in view of large density/viscosity ratios), 
one can restrict attention to the liquid film alone. The blowing/suction conditions at the wall $y=0$ are
\begin{equation}\label{eq:BC-wall}
u^* = 0, \qquad v^* = F^*(x^*,t^*),
\end{equation}
where $u^*,v^*$ are the streamwise (parallel to the inclined plane) and transverse (perpendicular to the plane) velocities, respectively, 
as depicted in Figure~\ref{fig1}.
The uncontrolled system admits a uniform flat film solution known as the 
Nusselt solution~\cite{kalliadasis2011falling}, given by $h^*(x,t) = h_0^*$ and a semi-parabolic in $y^*$ horizontal fluid velocity with
surface value $U^*_s = \frac{\rho^* g^* \sin \theta (h_0^*)^2}{2\mu^*}$, where $g^*$, $\rho^*$ and $\mu^*$ are the acceleration of gravity and the 
fluid's density and viscosity, respectively. An appropriate non-dimensionalisation of this problem allows us to define two important dimensionless 
parameters that characterise the system. The Reynolds number $Re = \frac{\rho^* U^*_s h_0^*}{\mu^*}$, and the 
capillary number $Ca = \frac{\mu^* U^*_s}{\gamma^*}$, which measure the relative importance between inertia and viscosity, 
and between gravity and surface tension (represented by $\gamma^*$), respectively.

As explained previously, full models such as the Navi- er--Stokes equations are computationally expensive to simulate. 
However, in the case of thin liquid films, the mean interface height $h_0^*$ is much smaller than the length of the domain, $L^* = nh^*_0$, so that we can define a 
\emph{long wave} parameter $\epsilon = \frac{h_0^*}{L^*} = n^{-1} \ll 1$. This disparity of scales facilitates a multiscale approach to derive
from first principles hierarchies of reduced-order models. For the remainder of the modelling discussion we use $h_0^*$ and $U^*_s$ as reference length and velocity scales 
alongside the defined groupings to transfer the system to its dimensionless counterpart (and drop the decorations accordingly). 
In this context, the requisite assumptions are:

\begin{itemize}
	\item[{\bf(A1)}] (long-wave assumption) the geometrical aspect ratio $ \epsilon $ is small;
	\item[{\bf(A2)}] The Reynolds number $Re$ is $\mathcal{O}(1)$;
	\item[{\bf(A3)}] Surface tension is sufficiently strong to appear at leading order, i.e., the capillary number is small, and $Ca = \mathcal{O}(\epsilon^2)$
	is the appropriate distinguished limit;
	\item[{\bf(A4)}] The controls $F$ are small $F = \mathcal{O}(\epsilon)$, implying weak blowing/suction.
\end{itemize}

Using assumptions {\bf(A1)}-{\bf(A4)} and asymptotic analysis techniques, Thompson \emph{et al.}~\cite{thompson2016stabilising} derived two different long-wave models. Both models 
satisfy a mass conservation equation
\begin{equation}
h_t + q_x = F(x,t),\label{eq:masscons}
\end{equation}
and couple with an equation for the flux $q(x,t)$.
In the first model, the Benney equation, they obtain an explicit expression for $q(x,t)$ and the model is a single PDE for the interfacial height $h(x,t)$:
\begin{align}
\label{eq:Benneyq}
q(x,t) =& \frac{h^3}{3}\left(2-2h_x\cot\theta +\frac{h_{xxx}}{Ca}\right) + \\ \nonumber
& +Re\left(\frac{8h^6h_x}{15}-\frac{2h^4 F}{3}\right).
\end{align}
The second model is the weighted residuals model, which describes the evolution of the interfacial height $h(x,t)$ and the flux $q(x,t)$:
\begin{align}
\label{eq:wrmq}
\frac{2 Re}{5}h^2 q_t + q =& \frac{h^3}{3}\left(2-2h_x\cot\theta +\frac{h_{xxx}}{Ca}\right) + \\ \nonumber
& +Re \left(\frac{18q^2h_x}{35} - \frac{34hqq_x}{35} + \frac{hqF}{5}\right).
\end{align}
We note that the controls appear as an inhomogeneous term $F(x,t)$ in the mass conservation equation \eqref{eq:masscons}, 
and this structure plays a crucial role in the efficiency of these controls.

Due to the asymptotic reduction, the models do not directly provide the evolution of bulk quantities such as the streamwise 
and transverse velocities $u$ and $v$, but these are known in terms of
the interfacial height $h(x,t)$ and the flux $q(x,t)$ from the analysis. At leading order, 
and using the results from the weighted residuals model, for example, the fluid velocities are
\begin{align}
u(x,y,t) &= \frac{3q}{h}\left(\frac{y}{h(x,t)}-\frac{y^2}{2h(x,t)^2}\right), \label{eq:uv1}\\ 
v(x,y,t) &= F(x,t) - \int_0^y u(x,y',t) \ dy',\label{eq:uv2}
\end{align}
thus providing a description of the flow field in the whole domain that can be compared with direct numerical simulations, for instance. 
Details on the numerical methodology behind solving these reduced-order models are provided in Appendix~\ref{B1:PDEsol}.
We underline however that we have used a unimodal perturbation of sufficiently small amplitude, typically of $\mathcal{O}(10^{-2})$, 
as initial interface shapes in all our numerical solutions, including higher up in the hierarchy at the DNS level to ensure consistency in the comparisons.

The above long-wave models are significantly more accessible computationally than the full Navier--Stokes equations, 
but they are still highly nonlinear 
and to-date have not been tackled analytically to obtain rigorous results. 
Hence, further simplifications are necessary in order to make analytical progress. 
One can, for example, perform weakly nonlinear analysis 
to obtain a Kuramoto--Sivashinsky (KS) equation for small but nonlinear perturbations from a flat interface~\cite{kalliadasis2011falling,thompson2016stabilising}. 
The KS equation is a fourth order nonlinear PDE having the same form as~\eqref{e:feedback_PDE} 
with $\mathcal{L}u = u_{xxxx} + u_{xx}$ and $\mathcal{N}(u) = uu_x$. The KS  equation appears  in a plethora of applications and is widely studied since it is one of the simplest model PDEs which exhibit spatiotemporal chaotic behaviour. Over the last few decades existence and uniqueness of solutions has been explored~\cite{tadmor1986well}, different types of attractors have been characterized~\cite{collet1993global}, and the route to chaos for solutions of the KS equation have been reported~\cite{papageorgiou1991route}, thus exemplifying the range of interesting analytical 
and computational results that can be achieved even at this lowest member of the model hierarchy. 

\begin{figure*}
\centering
\includegraphics[width=0.99\textwidth]{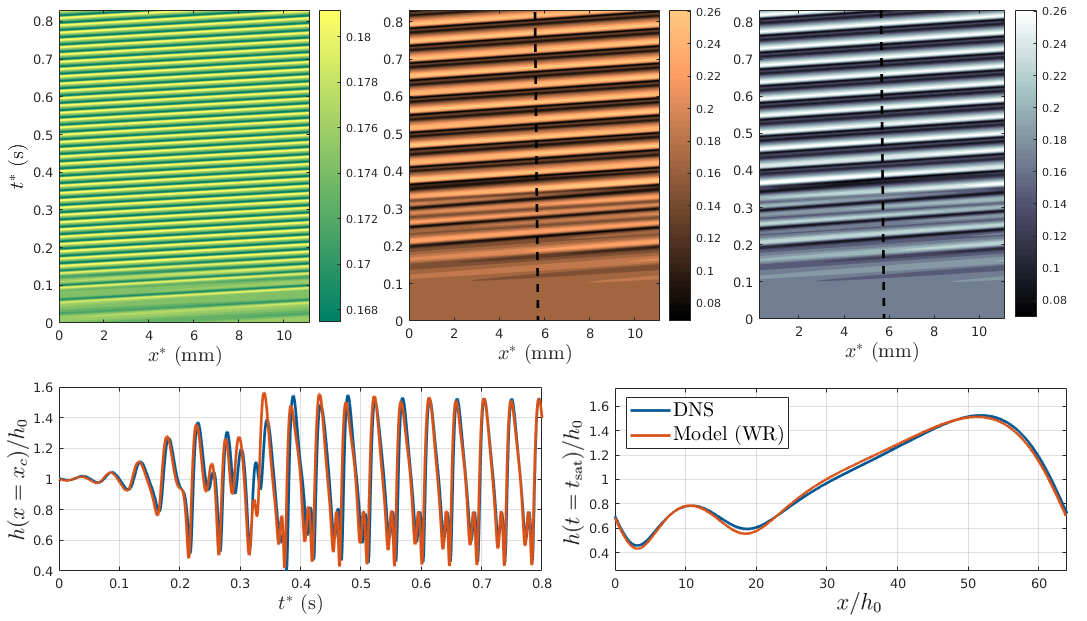}
\caption{Validation summary for a film of undisturbed thickness $h_0^*=175\ \mu$m on an inclined plane described by $\theta = \pi/3$. 
The top panel, shows, from left to right, the interfacial film height $h^*$ as obtained from calculations based on the Kuramoto--Sivanshinsky equation, 
the weighted residuals long-wave model, as well as direct numerical simulations. Dashed lines indicate a virtual measurement station for the interfacial height 
showing the evolution of the film in the bottom left panel. The bottom right panel compares saturated interfacial shapes for the long-wave model and the DNS.}
\label{fig:validation}
\end{figure*}

Some uncontrolled computations are presented next to showcase the underlying nonlinear dynamics.
Figure~\ref{fig:validation} compares the results for the evolution of the interfacial height using the KS equation (top-left panel), 
our most comprehensive long-wave model (the weighted residuals model, top-middle panel) and the virtual experiment solving the Navier--Stokes equations (top-right panel) 
for a film thickness of $h^*_0 = 175\ \mu$m and an inclination angle of $\theta = \pi/3$.
These $x-t$ plots are colour coded according to the film thickness with darker colours representing thinner regions.
In all cases after a short transient a nonlinear travelling wave emerges moving from left to right as seen from the straight
lines in the $x-t$ plane followed by wave troughs and crests.
The bottom left panel presents the evolution of the film thickness measured at a fixed station positioned at the centre of the $L^*=64h^*_0$-sized periodic 
domain used for the computations, highlighted by a black dashed line in the figure. The time-periodic signal (after an initial transient)
once again verifies the presence of a nonlinear travelling wave of permanent form. Finally, the bottom right panel 
presents a comparison between the saturated interfacial profiles of the resulting nonlinear wave, as obtained from both long-wave model and DNS predictions.
Notably, from both the transient and the final state comparisons, 
there is excellent agreement between the long-wave model and the DNS - this is the case as long as the assumptions 
made in the derivation of the model are used for the DNS.
This is in fact a stringent scenario for all the reduced-order models, as inertial effects and nonlinear features lead to specific forms of breakdown (which we will soon describe) 
in this region of the parameter space. However a weighted-residuals approach is still reliable at this stage.
By contrast, a setting weighed down by restrictive assumptions as for the KS equation, 
leads to quantitatively different solutions unless the resulting dynamical
behaviour is simple; this is illustrated by the convergence of the solution to the KS equation to a bimodal travelling wave, instead of the unimodal wave 
that both the weighted residuals and the DNS converge to, see bottom-right panel of figure~\ref{fig:validation}. 

Given these results, one could question the appropriateness of some of the reduced-order models in direct comparisons with
DNS and experiments (of course
the dynamics supported are rich and the equations are of fundamental mathematical importance). Notably,
both the Benney and the KS equations are valid in very limited parameter 
regimes, making comparisons with experiments difficult. 
Even the weighted-residuals model solutions included in the comparisons with DNS in figure~\ref{fig:validation} are
close to the boundary of their applicability, and numerical solution is already hindered by stiffness.

Simpler mathematical models, however, are key players in mathematical studies 
and help us to push conceptual boundaries to the point where the developed methodologies can be applied higher up in the model hierarchy.
This is the approach taken here and in particular we subsequently use such methodologies 
in our virtual experiments with the aim of utilizing them in real-world scenarios, e.g. physical experiments and applications. 
We should point out that small discrepancies still exist 
between any model and the full DNS. As illustrated, the long-wave model and DNS are in quantitative agreement (even during
transient dynamics towards equilibrium coherent structures), hence we are confident that comparisons and hybrid use of the two frameworks
across a wide range of scenarios are appropriate.

More recently, there has been interest in the study of feedback control for the Kuramoto--Sivashinsky equation. Armaou and Christofides~\cite{armaou2000feedback} explored the control of the zero solution (flat state) in small domains, while Gomes \emph{et al.}~\cite{gomes2015controlling,gomes2017stabilizing} 
generalised their results to long domains (where chaotic behaviour is observed) and to stabilising solutions with a chosen
non-uniform interfacial shape. 
The results in~\cite{gomes2015controlling,gomes2017stabilizing} show that any possible solution to the KS equation can be stabilised using a finite number of point actuated controls 
whose strength only depends on the difference between the observed and desired interfacial shapes. 
The number of control actuators depends only on the domain length, and the control rule can be computed using a standard pole placement algorithm~\cite{kautsky1985robust}. 
Furthermore, the controls are robust to uncertainty in the problem parameters, as well as to small changes in the number of controls used.
Motivated by the similar linear stability properties between the KS equation and the Benney equation (the simplest long-wave model), 
Thompson \emph{et al.}~\cite{thompson2016stabilising} studied the control problem for two long-wave models: the Benney equation and 
the first order weighted residual model, which acted as a test for the robustness of the controls across the full hierarchy of models. 
The authors start by showing that in the unrealistic scenario where one can observe the whole interface and \emph{actuate everywhere}, 
the simplest proportional controls of the form 
\begin{equation}
\label{e:full_prop}
f(x,t) = -\alpha(h(x,t)-1),
\end{equation}
for some constant $\alpha>0$ to be determined, efficiently drive the system towards the flat solution $h(x,t) = 1$ (or indeed any desired solution $H(x,t)$, 
by replacing $1$ by $H(x,t)$). The critical value of $\alpha$ can be computed from linear stability analysis of the Benney or KS equations 
and it depends only on the Reynolds and capillary numbers. 
It is also shown that the critical $\alpha$ for the Benney equation is sufficient to obtain linear stability of the weighted residuals model 
and indeed the full Navier--Stokes equations, by solving an Orr--Sommerfeld system. The critical value of $\alpha$ for the Benney equation is calculated by solving a linear algebraic equation for the eigenvalues of the linearized system, while for the weighted residuals model the associated equation is quadratic. For the full model, the resulting Orr--Sommerfeld equation is a fourth order differential equation for each eigenvalue, which needs to be solved numerically, as was done, e.g., in \cite[Section III-C]{thompson2016stabilising}. Even though the result for the Benney equation underestimates the boundaries of linear stability, leading to stronger controls (larger $\alpha$) than necessary, the relative simplicity and reduced computational cost offered by the Benney equation is a clear advantage of using reduced-order models. 
Nonlinear stability is confirmed by numerical simulations of the initial value problem.

In the more realistic case of a finite number of observations of the interface and a finite number
of control actuators, Thompson \emph{et al.}~\cite{thompson2016stabilising}  use proportional feedback controls of the form
\begin{equation}
\label{e:PA_controls}
f(x,t) = -\alpha\sum_{j=1}^M \delta(x-x_j) (h(x_j-\phi,t)-1),
\end{equation}
where $\delta(\cdot)$ is the Dirac delta function, the control actuators are located at the positions $x_j, \, j=1,\dots,M,$ and 
observations of the interface are made at $x=x_j-\phi$ for some displacement $\phi$. Such a control protocol was shown to be efficient 
in stabilising the flat solution $h(x,t) = 1$ for $M$ sufficiently large (in practice, $M=5$ is usually sufficient 
for Reynolds and capillary numbers found in 
relevant flows -- see Appendix~\ref{A:parameters}). 
The authors explore other forms of controls, such as the case when the whole interface is observed (in which case one can use pole placement algorithms, 
similarly to the ones used for the KS equation), or when the number of observations and controls are different, 
in which case these can also be combined using \emph{dynamic observers}~\cite{thompson2016stabilising,zabczyk1992mathematical}. 

The latter control strategies are the most efficient in stabilising the flat solution for the Benney equation, but since their design is model-dependent, 
their applicability across the hierarchy of models is unlikely to be accurate. For this reason, in this paper we chose to use DNS to study the applicability 
of \emph{full proportional controls}~\eqref{e:full_prop} and \emph{point-actuated proportional controls}~\eqref{e:PA_controls} developed for the long-wave models.
We will see that we cannot simply ``translate" the controls designed for the long wave models (even the weighted residuals model) into the virtual experiment 
framework, since there are physical effects that appear at the DNS level which are not mitigated in the weighted residuals model. 
However, our study enables us to design a simple adaptation of the model-based control rules to attain desired control
in the DNS, thus acting as a valuable guiding tool within the multi-dimensional parameter space and reducing computational time requirements by several orders of magnitude.

We point out that an alternative to proportional feedback controls would be to use an optimal control approach \cite{troltzsch2010optimal}, where we would instead minimise a cost functional constrained by the PDEs which constitute our hierarchy of models. However, this approach is much more computationally expensive, as it involves a gradient descent algorithm that requires solving the highly nonlinear PDEs and corresponding adjoint equations multiple times \cite{borzi2011computational}, and not necessarily achievable theoretically, as functional analytical tools required to prove existence of an optimal control \cite{troltzsch2010optimal} are not available in these highly nonlinear settings. Furthermore, as we will discuss in Section~\ref{s:numerics}, one of the strengths of the approach under consideration in the present context is that controls based on linear stability analysis of the reduced-order models will be sufficient in order to stabilise the full nonlinear system.

\subsection{DNS solution of the Navier--Stokes equations}

We have constructed a state-of-the-art computational framework in which we can conduct highly accurate \textit{in silico} experiments
of a real-world scenario, controlling a falling film down an inclined plane. This framework does not require any restrictive assumptions 
and is capable of resolving all the relevant scales and nonlinearities, thus enabling direct comparisons with real-world physical experiments,
and indeed theoretical predictions that are devoid of experimental errors and challenges in imaging, measurement and data acquisition.
As such, it constitutes a powerful environment to evaluate the 	mathematical model prior to refining the control methodology for specific applications. 
In addition, we can construct databases containing the entirety of the flow information without 
experimental restrictions and errors. Importantly, everything above holds in a general setting;
however, we selected a classical fluid problem for which a range of well-known hierarchy of reduced models, 
numerical and experimental results are available.
This allows us to focus on the most delicate aspect of our work, namely the efficiency and accuracy of controls constructed on reduced models
as they are used in the control of progressively more complex systems, e.g. Navier-Stokes DNS.
Furthermore, the fluid flow problem is of intrinsic importance and offers a rich landscape of solutions and interplay of physical effects 
and pertains to a wide range of industrial applications from coating technologies to cooling systems in microchips and multi-physics 
solutions for heat and mass transfer. 

Details of the open-source computational platform we used, the $\mathpzc{Gerris}$ Flow Solver \cite{popinet1,popinet2},
are provided in Appendix~\ref{B:DNSdetails}. There we discuss the general direct numerical simulation methodology, details about the discretisation 
scheme, the volume-of-fluid method used to represent the fluid interface, as well as technical aspects related to the large scale solution effort, 
data gathering and post-processing. A rigorous validation procedure has also been implemented utilizing both converged and transient model solutions,
and also information regarding stability and regime boundaries in the target parameter space (discussed in detail in the following section).

The implementation is based upon monitoring a periodic domain of sufficient length compared to the film height (typically $64$ times the film thickness)
while being able to investigate all flow quantities of relevance in an unsimplified setting. This still enables us to import the full control toolkit
presented at the end of subsection~\ref{ss:ROM}, including either distributed (full-surface) or, more realistically, point-actuated controls 
based on discrete interfacial height measurements.

Distributed controls are approximated as piece-wise constant strips -- numbering between $4$ and $64$ across the length of the periodic domain -- attempting to mirror 
a realistic setup with changeable parts and modular elements. We have noticed that beyond $16$ elements the results no longer vary within the tested conditions,
thus amounting to a sufficiently accurate representation of the full continous setup.

As in Thompson et al. \cite{thompson2016stabilising}, for the point-actuated scenario we require a special treatment of the localised control region.
The Dirac $\delta(x)$-functions introduced in equation~\eqref{e:PA_controls}
are converted to smooth finite counterparts via $s(x) = \exp [ (\cos(2\pi x) - 1)/w^2 ]$, where $w$ denotes the smoothing window. As $w \to 0$,
we find $s(x) \to \delta(x)$, but in practice we choose $w^2$ to be of $\mathcal{O}(10^{-3} - 10^{-2})$ to preserve the nature of the intended effect 
while allowing for an efficient numerical solution of the resulting system given resolution and multi-scale constraints.


\section{Results}
\label{s:numerics}

\begin{figure*}[!h]
\centering
\includegraphics[width=0.7\textwidth]{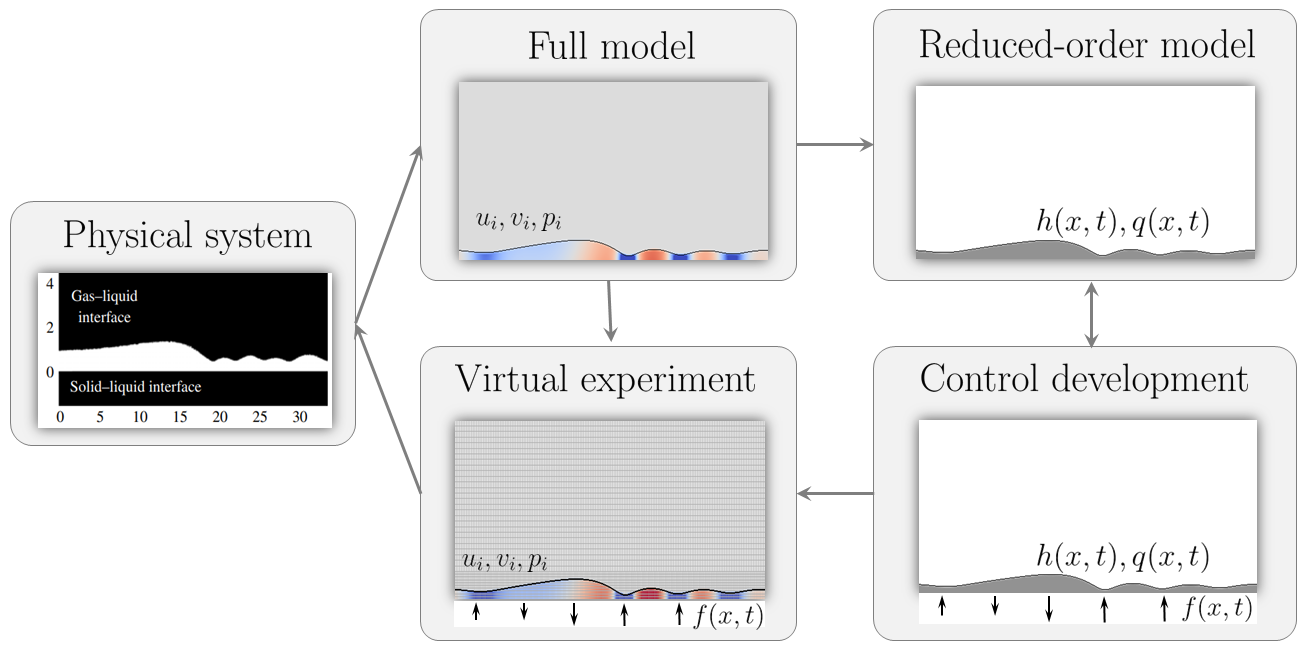}
\caption{Multi-layer control methodology and connectivity outline, from experimental study (\cite{denner2018solitary}, adapted) and application design in the context of falling films down inclined planes  to the associated reduced-order models and \textit{in silico} experiments.}
\label{fig2}
\end{figure*}

Here we present our results of controlling a falling liquid film in a DNS framework, as informed by control-theoretical approaches for reduced-order models. 
Figure~\ref{fig2} encapsulates our multi-stage approach to the control problem, where our fundamental aim is to be able to accurately and efficiently 
control the dynamics of a complex physical setup. Trial and error in such a delicate and expensive environment is inefficient,
hence modelling 
and simulation-based approaches are to be used in conjunction to inform a cogent strategy. Full model representations of the target system still introduce
complexity and efficiency barriers, often offering no advantages apart from shifting the difficulty and burden to
expensive computational clusters.
Our approach is to use a variety of reduced-order models to construct controls and use the 
obtained knowledge to pinpoint controllable regimes 
and guide the heavier simulation machinery to pre-determined regions in parameter space.
The computational 
platform can then be expanded to overcome modelling assumptions and employ the same control strategies in a targeted way, emulating challenging 
regimes in the real-world physical system itself.

For our particular model choice of multi-phase system, water and air at room temperature are chosen as reference. 
Other fluids could be used (see Appendix~\ref{A:parameters} for a brief discussion), 
however we have opted for a more challenging yet realistic scenario that supports a balance of competing forces in the flow including 
inertia that is typically absent in an oil-based flow.
The controls are built based on linear stability analysis 
and the numerical solution of a first-order weighted residuals model whose suitability has been discussed in subsection~\ref{ss:ROM}. 
Similar results can be obtained using different reduced-order models, but with notable restrictions in terms of applicability in the parameter space.

Given the surprisingly good agreement between simulations of the reduced-order models and DNS 
(see Appendix~\ref{B:DNSdetails} and Figure~\ref{fig:validation}), we would perhaps expect to be able to use the controls developed for the model 
directly in the simulation of the full Navier--Stokes equations. 
If this were to work, the methodology would require no complicated data measurements (e.g., interfacial height) which would be needed in real time, 
and the control could be ``pre-coded", fast, and efficient. This is what we call ``offline" controls. 
Alternatively, if unsuccessful, one would need to use feedback controls, which are built using linear stability information from the reduced-order models, 
but which require readings from the full system. We explore each of these approaches next.

\subsection{Offline controls}
\label{ss:offline}

\begin{figure*}[h!]
\centering
\includegraphics[width=0.9\textwidth]{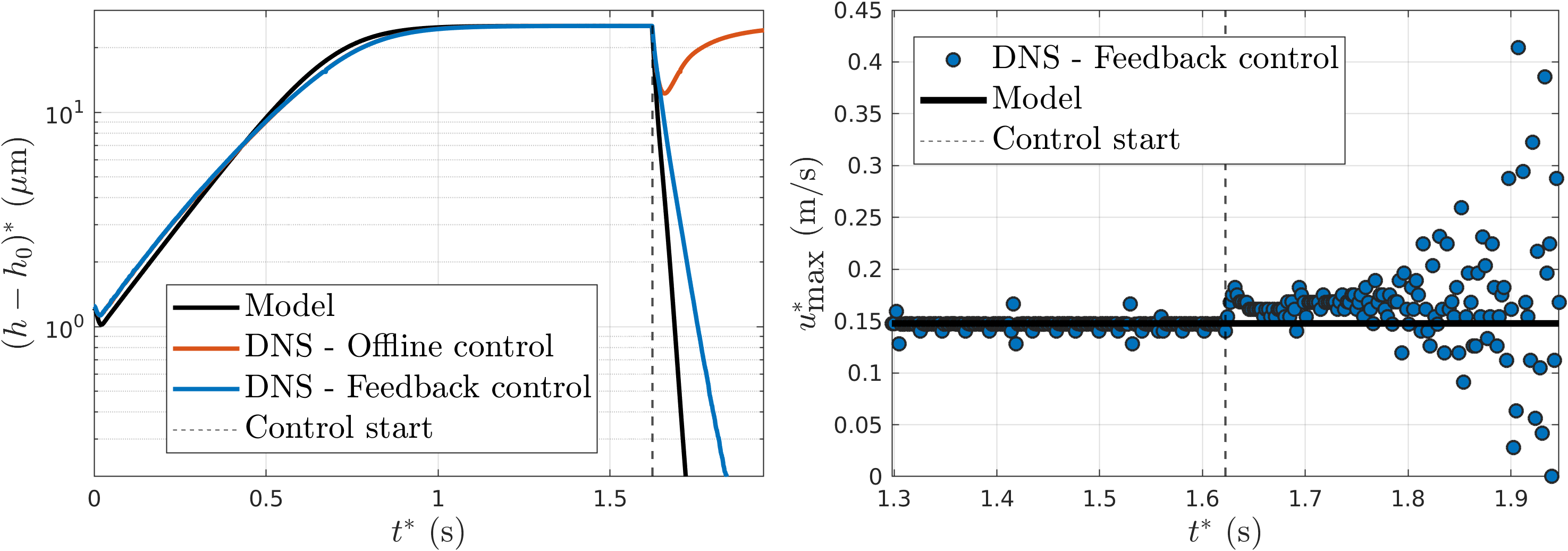}
\caption{Control results for a film of undisturbed thickness $h_0^*=125\ \mu$m on an inclined plane described by $\theta = \pi/3$. 
(Left) Evolution of the interfacial shape maximum $h^* - h_0^*$ for model and two DNS control approaches (offline and feedback).
(Right) Measured instantaneous velocity based on the interface maximum location for both model (black line, constant) and DNS (symbols) for the 
feedback control case. In both panels the start time of the controls ($t^* \approx 1.62$ s) is shown via a dashed vertical line.}
\label{fig:offline}
\end{figure*}

It is tempting, given the potential for complexity reduction, to conduct a full control feasibility study at the level of the reduced-order model and 
then simply export the resulting controls into the DNS. 
Should this work, all values of interest could be stored and relatively easily adapted to flow conditions at the level of the full setup.
The initial motivation for this line of investigation followed an interesting observation regarding model behaviour; when the control 
actuation is of the form~\eqref{e:full_prop}, 
after a very short-lived initial adjustment, the dynamics is steered towards a linear regime and for the rest of the evolution towards a flat state,
the interface decays in amplitude and translates with a constant speed, keeping the same shape to within a translation and lateral rescaling.
Furthermore, the decay in time of its amplitude is exponential since the controlled dynamics are stable and linear. In fact, for the Kuramoto-Sivashinsky equation which is the simplest model for falling liquid films presented in section~\ref{ss:ROM}, it is possible to prove that this decay is exponential and to estimate its decay rate, see~\cite{gomes2017stabilizing}.
Thus it would seem reasonable that, using normalisation and spatial readjustment (left/right shifts), the precise interfacial 
profiles would inform a very simple setup for the proportional control functional $f(x,t)$ as defined in eq.~\eqref{e:full_prop}.
This can be summarised as a separation of variables via $f(x,t) = g(x - c t) \cdot \exp(-\omega t)$, where the shape $g$ would be given 
by the nonlinear saturated wave shape moving with a characteristic constant velocity $c$, 
while the time-dependent amplitude (informed by decay rate $\omega$) could be readily extracted from the model. 
Making sure of the alignment, i.e. starting with a suitable shift in both space and time, would be the only sensitive aspect of the procedure, 
since it does require a reasonable level of synchronisation. A typical result of this procedure is illustrated in red as part of Figure~\ref{fig:offline} 
for a relatively mild case given by a liquid film of height $h_0^*=125\ \mu$m falling down an plane inclined at an angle of $\theta = \pi/3$. 

In this numerical experiment, the control is applied at $t\approx 1.62$ s (this time is indicated by a vertical dashed line), 
after convergence to a nonlinear saturated shape. The initial evolution looks promising, with a rapid decay of the perturbation being observed. This occurs, however, only during a very brief window 
and is followed by an equally rapid return to the nonlinear wave shape we were aiming to control to begin with (though this is may not necessarily be the case 
and a convergence to a different solution is also possible in principle). Albeit a specific example, this behaviour is characteristic 
of every case study attempted with this strategy. The action of the imposed control eventually falls out of sync with the interface 
due to nonlinearities and the richer physics in the full model. During the respective time, the imposed control continues to decay exponentially and its strength eventually becomes insufficient to affect the flow, which reverts to an uncontrolled configuration. Further tailoring of the imposed control parameters (in terms of velocity, amplitude etc.) results in variations around this trend, and without a perfect co-evolution, the precise self-similar interfacial structure is not preserved, resulting in a strategy which is not robust and ultimately fails. 

Further details are best explained in contrast to full proportional control, in which we follow and use the interface shape at every timestep
to inform the construction of $f(x,t)$ rather than its modelled offline counterpart. 
The results in question are shown in blue in Figure~\ref{fig:offline}. The panel on the left provides a 
useful indication that in this case, the decay of the interfacial perturbation is indeed still exponential in time, however at a slower rate 
than in the reduced-order model. But arguably even more revealing is a study of the horizontal velocity of the interface as a function of time (right panel). 
Even in a successful control scenario, the variation once the control is applied is significant and becomes progressively more sensitive 
as the perturbation decays in time. There should thus be no surprise that a fixed shape, along with its velocity and a decay rate would be near impossible 
to determine reliably in a practical manner, even with perfect information. Any such fixed parameters would rapidly lead to an out-of-phase behaviour, 
with the exponentially decaying control ultimately becoming null while the system state is still considerably perturbed. Whilst unsuccessful, this study has given us valuable insight in terms of the difference in behaviour between the reduced and full models. 
It has also provided a strong early indication that, while offline controls can only achieve a transient positive response, if the capabilities were enhanced towards the integration of dynamic observations via feedback controls, 
the desired convergence towards a target state can be recovered. We explore this further in the following subsection.

\subsection{Feedback controls}
\label{ss:feedback}

The findings of the previous section indicate that time-dependent flow information needs to be incorporated into
the control methodology.
Broadly speaking, there are two types of feedback control we can use: 1. distributed controls, which can act in principle on the 
entire bottom surface of the system  and are given by~\eqref{e:full_prop}, and 2. point-actuated controls given by~\eqref{e:PA_controls}, which are only defined at discrete points along the control surface.
Both of these can rely on either complete or limited information on the interfacial shape. There is, however, a clear difference 
in applicability between the two.  The former is more idealized in view of the need of full information for the interfacial shape and
continuous actuation on the boundary.
On the other hand, a practical setting will only offer the possibility 
of limited flow information (interfacial shape, velocities, pressure) and a specific range of
actuation points are feasible
in order to preserve the structural integrity of the boundary. 
Both scenarios are explored in what follows. 

For the DNS implementation each of the above methods requires a suitable discretization treatment.
For distributed controls, we have opted to construct a full-surface actuation method based on equally-sized 
segments that would represent actuation strips on the bottom boundary. For our target parameter regimes, we have noticed results no longer 
vary beyond $16$ such segments, however in most cases $64$ have been used in order to ensure smoothness even in more challenging nonlinear scenarios.
In the case of point-actuated controls, the theoretical $\delta-$function profiles were treated 
following Thompson et al.~\cite{thompson2016stabilising}, using smoothed actuator shape functions
(with weights $w=\mathcal{O}(10^{-3} - 10^{-2})$) extending 
over several computational grid cells in order to suitably incorporate the required variation, whilst maintaining the total flux of fluid in 
the domain constant via an integral normalisation.

With the above computational framework in place, we need to fix an observation strategy that is both powerful and practically achievable, with the 
interfacial height location proving sufficient to satisfy the required criteria. This is sufficient to fully inform the reduced-order models, 
while obtaining and processing additional information regarding the flow field on the fly would be too expensive (with currently available technology) to 
allow for a timely control adaptation. Assuming for simplicity that a full interfacial shape approximation $h(x,t)$ is available (this need not be the case, and limited observers 
are also possible, see~\cite{armaou2000feedback,thompson2016stabilising}), the control procedure can be summarised using the following parameters:
\begin{enumerate}
 \item control strength $\alpha$, typically a multiplier of the difference in absolute value between the current and the target states;
 \item an upstream or downstream displacement $\phi$, allowing for the possibility of incorporating spatial delay information in the actuation procedure;
 \item for the point-actuated setup, the number of application points $M$ -- in this case, we also consider a limited number of observations, 
 equal to the number of actuators.
\end{enumerate}

We note that, for the point actuated case, a more complicated (and possibly more efficient) control strategy can 
be obtained with observations of the full interface. This would involve using the full interface information to design an optimal 
feedback control strategy, which can be achieved using pole placement or a linear quadratic regulator -- this was done in~\cite{thompson2016stabilising}, where a feedback control strategy was designed for the Benney equation and tested in the 
weighted residuals model. Furthermore, if the number of observations is different than the number of actuators (or if a distributed control were to be designed using a finite number of observations of the interface), similar strategies can be used to design dynamic observers (see~\cite{armaou2000feedback,thompson2016stabilising}). Both of these strategies are the subject of current work, as they are harder to translate across the hierarchy of models and are therefore beyond the scope of this paper.

\subsubsection{Controlling to a uniform state}

In the distributed control case, for given parameters of the system (such as the Reynolds number $Re$, capillary number $Ca$ and the inclination angle $\theta$), 
we can perform linear stability analysis of the simplest reduced-order model -- the Benney equation -- and obtain an analytical expression for the minimum 
value of $\alpha$ needed to obtain stability; it was shown in~\cite{thompson2016stabilising} that these values are sufficient to guarantee linear stability 
of the full system. In contrast, if the controls are point actuated, the eigenvalues of the linearised system need to be computed numerically. However, 
the reduced-order models still provide valuable information in this case, such as an estimate on the number of control actuators sufficient to obtain stability. 
For the Kuramoto--Sivashinsky equation this is shown to be larger than or equal to the number of unstable Fourier modes, 
see~\cite{gomes2015controlling,gomes2017stabilizing}, and this was observed to be sufficient for both long wave models in \cite{thompson2016stabilising}. 

\begin{figure*}[h!]
	\centering
	\includegraphics[width=0.43\textwidth]{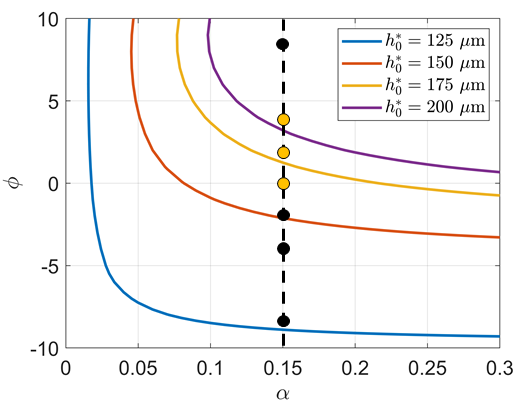}\,
	\includegraphics[width=0.55\textwidth]{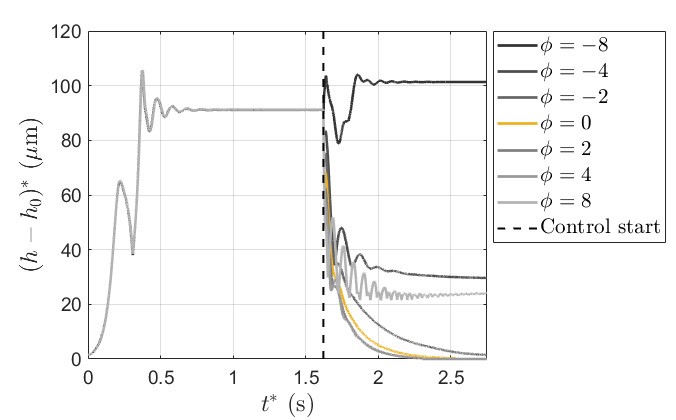}
	\caption{Distributed control study summary for a test case described by 
		films of undisturbed thickness $h_0^* = 125\ \mu$m-$200\ \mu$m falling down a plane with inclination angle $\theta = \pi/3$, steered towards a flat film. 
		(Left) Linear stability results as predicted by the weighted-residuals reduced-order model and associated 
		direct numerical simulation results for control strength $\alpha = 0.15$ and undisturbed interface height $h_0^* = 175\ \mu$m. 
		The regions above each curve denote areas of expected linear stability.
		Coloured symbols indicate a successful control scenario, 
		while black circles denote convergence to a non-flat state in the DNS. 
		(Right) Evolution of interfacial maxima as a function of time 
		for the family of tests with fixed $h_0^* = 175 \mu m$, $\alpha = 0.15$ and different shifted observer values $-8 \le \phi \le 8$.
		}
	\label{fig:fullsurface}
\end{figure*}

\begin{figure*}[!h]
\centering
\includegraphics[width=0.43\textwidth]{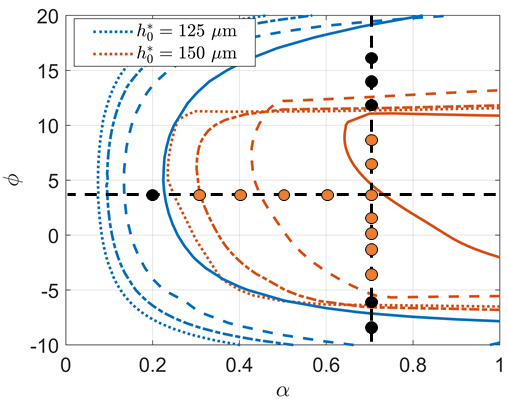}\,
\includegraphics[width=0.55\textwidth]{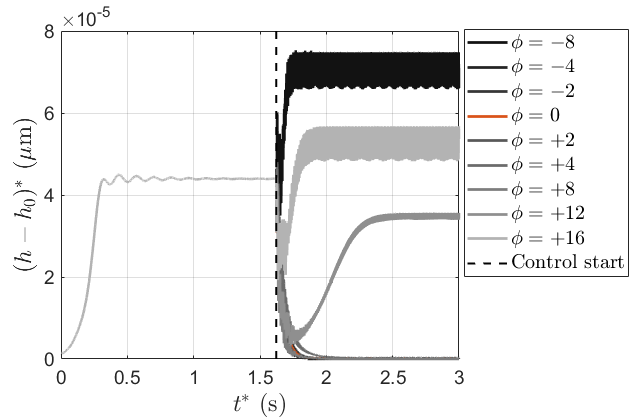}
\caption{Point-actuated control study summary for a test case described by 
films of undisturbed thickness $h_0^* = 125\ \mu$m-$150\ \mu$m falling down a plane with inclination angle $\theta = \pi/3$, steered towards a flat film.
(Left) Linear stability results as predicted by the weighted-residuals reduced-order model
with different numbers of control application points: $M=3$ (solid), $M=5$ (dashed), $M=7$ (dash-dotted) and $M=9$ (dotted),
and associated 
direct numerical simulation results with $M=5$ controls for fixed $h_0^* = 150\ \mu$m and either control strength $\alpha = 0.7$ and varying $\phi$, 
or fixed shifted observer value $\phi=4$ and varying $\alpha$. 
The regions bounded by each curve denote areas of expected linear stability.
Coloured symbols indicate a successful control scenario, 
while black circles denote convergence to a non-flat state in the DNS. 
(Right) Evolution of interfacial maxima as a function of time 
for this family of tests with undisturbed thickness $h_0^* = 150\mu m$, $M=5$ controls, fixed control strength $\alpha = 0.7$ and 
different shifted observer values $-8 \le \phi \le 16$.}
\label{fig:pointactuated}
\end{figure*}

\begin{figure*}[!ht]
\centering
\includegraphics[width=0.99\textwidth]{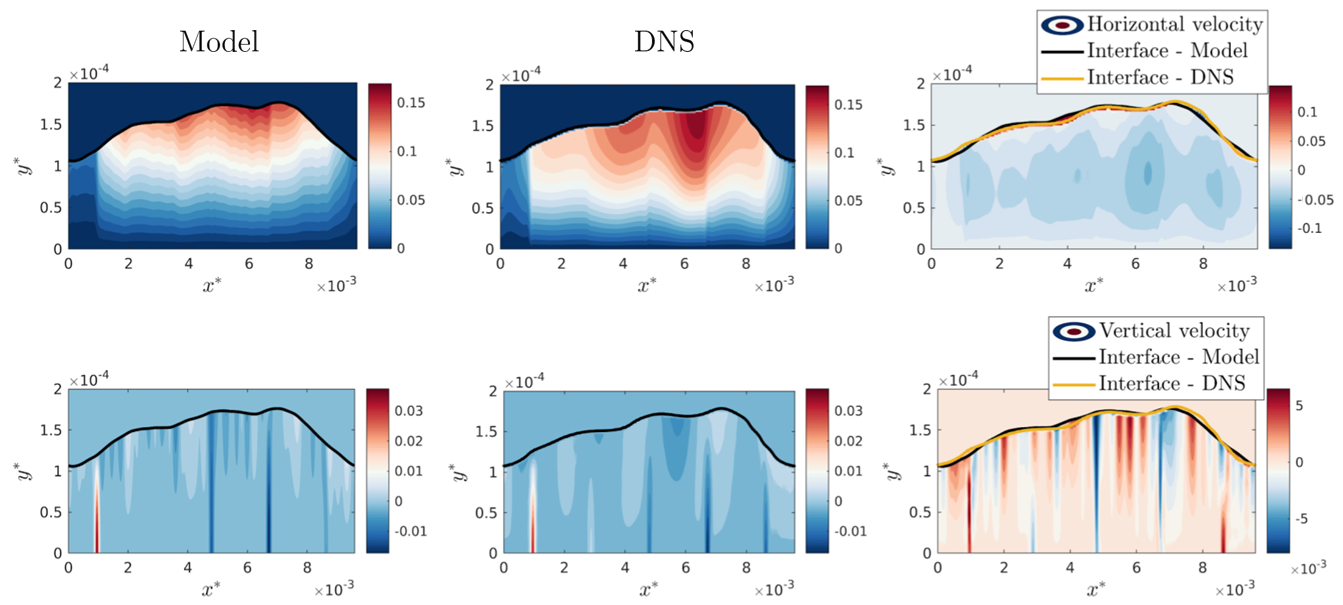}
\caption{
Comparison between horizontal (top row) and vertical (bottom row) velocity fields for both reduced-order model (left panels) 
and direct numerical simulation (central panels) results, with their 
difference in absolute value shown on the right panels. The snapshots are taken $0.01$ s after the application of a point actuated control of strength $\alpha=1.0$
with a shift of $\phi = 4$
to the developed solution of a liquid film of undisturbed thickness $h_0^* = 150\ \mu$m falling down a plane with inclination angle $\theta = \pi/3$. 
A summary animation of the post-control dynamics is available as supplementary video material.}
\label{fig:maxComparison}
\end{figure*}

We perform linear analysis for both control cases in Figures~\ref{fig:fullsurface} and~\ref{fig:pointactuated}. 
Figure~\ref{fig:fullsurface} presents the results for distributed controls; in the left panel we plot the linear stability regions predicted by the first order weighted-residuals model for films with undisturbed thicknesses spanning $h_0^* = 125\mu$m -- $200\mu $m and inclination angle $\theta = \pi/3$. These regions were obtained by computing the eigenvalues of the linearised system by including the effects of the spatial delay $\phi$ in the analysis in~\cite[Sections III.B,E]{thompson2016stabilising}, and plotting neutral curves determined by the maximum real part of the resulting eigenvalues being zero. We note that due to the longwave assumptions, we only considered eigenvalues corresponding to long-wave modes for the analytical computation. However, since linear stability does not guarantee that the nonlinear evolution of the controlled system is stabilised, we also performed a number of tests whereby we fix the undisturbed thickness $h_0^* = 175\ \mu $m and the control strength  $\alpha = 0.15$, and vary the shifted observer values by $-8 \le \phi \le 8$, corresponding to the dashed line in the left panel. The evolution of the interfacial maxima as a function of time for this family of tests is plotted in the right panel, and the plot also includes the subsequent dynamics after the controls
are switched on at $t=1.6$ s indicated by a dashed line.
Figure~\ref{fig:pointactuated} was obtained in a similar way, but here the eigenvalues of the linearised system were obtained numerically using the Jacobian of the system. The left panel shows the stability regions (defined as above) for $h_0^* = 125 \mu$m and $150\mu$m, and for varying $\alpha, \, \phi$ and $M$, the number of actuators (and observers). For this case, we tested the nonlinear evolution for $h_0^* = 150\mu$m, $M = 5$, and various values of $\alpha$, for fixed $\phi = 4$, and of $\phi$ for fixed $\alpha = 0.7$. The right panel plots the evolution of the interfacial maxima as a function of time for fixed $\alpha$ and varying $\phi$, where again we
include the dynamics after the controls are switched on indicating that the value of $\phi$ is crucial in attaining a controlled flat state. 

As pointed out earlier, prediction of linear stability (or instability) does not guarantee that the time evolution of the nonlinear system 
results in convergence to a flat state.
Indeed we observe that in the distributed case presented in figure \ref{fig:fullsurface}, 
the test described by $h_0^* = 175\ \mu$m, $\alpha = 0.15$ and $\phi = 0$ is predicted to be linearly unstable 
as it sits outside the corresponding neutral curve depicted in orange, i.e., it has at least one eigenvalue with a positive real part. However, the time evolution shows that the controlled interface evolves towards a flat state. On the other hand, for the same film thickness $h_0^* = 175\ \mu$m, 
control strength $\alpha = 0.15$, but when the observation shift is $\phi = 8$, linear stability predicts that controls stabilise the flat solution, 
however the controlled dynamics evolves to a
non-uniform equilibrium solution instead of the flat target state.  In the latter case, we found (not shown) that the dynamics evolves to a trimodal wave, which corresponds to an unstable short-wave mode in the reduced-order model that was not considered in the stability plots due to its short-wave nature. This short wave destabilisation is most likely triggered by the fact that $\phi = 8$ is a large shift, and hence information propagates from the wrong place. 
Similarly, in figure~\ref{fig:pointactuated} we observe that in the tests for the point actuated case with fixed $h_0^* = 150\ \mu$m and $M = 5$ controls, there are two values for $\alpha$, for fixed $\phi = 4$, for which linear stability analysis predicts the flat state to be unstable to perturbations, but for which the nonlinear simulations show that the flat solution is stabilised. Furthermore, there is one value of $\phi$ for fixed $\alpha = 0.7$ which is predicted to evolve to a flat solution while the DNS shows that the solution evolves  to a non-uniform shape. However, all of these values are close to the boundaries of stability, and it is therefore not surprising that this behaviour is observed.

Having described the general behaviour and accuracy of our models, we proceed with the systematic differences observed between the reduced-order models and the DNS. We have used a typical 
case, moderate film thickness $h_0^* = 150\ \mu$m, and reasonably strong control strength $\alpha = 1.0$ with $M=5$ application points 
and a shift of $\phi=4$ and calculated the discrepancies arising from blowing/suction controls as we move up the hierarchy of nonlinearities.
Figure~\ref{fig:maxComparison} presents the horizontal velocity fields (top panels) and the corresponding
vertical velocity fields (bottom panels) for the model and DNS in the left and middle panels as labelled, 
along with the difference between the two velocity fields (right panels) at $0.01$ s after the control is initialised.
The corresponding time evolution, showing additional quantitative information, is also included as supplementary video material.
The horizontal velocity fields are reasonably close to each other, both qualitatively and quantitatively.
For the vertical velocity fields, looking beyond the general qualitative and quantitative agreement, the most poignant feature
is the stronger vertical banding of the observed structures in the model. The transfer of information from the blowing/suction imposed 
at the bottom boundary reaches the liquid-gas interfaces more rapidly, while in the case of the DNS a dampening of this effect is observed 
as we move laterally. 
These discrepancies originate in the simplifying assumptions that were necessary in the derivation of the reduced-order
model but not in the DNS. These include, (a) depth-averaging, (b) neglect of higher-order effects leading to what is known as viscous dispersion
in falling film flows~\cite{ruyer2000improved}, and (c) inertial effects that are fully present in the DNS, which also make a difference 
given the considered moderate Reynolds number regimes.
Interestingly, the most discernible differences occur at the center of the fluid region rather than near the interface, 
and generally take a much less localised
form. Detailed comparisons between model and DNS for
particular cases provides insight into the action of the control mechanism 
and helps identify predictive bottlenecks, e.g. by overplaying the control effects in the case of thicker films in which DNS indicates that 
diffusion and stronger inertial features are likely to damp the intended control strength, resulting either in a delayed convergence to a steady state or 
insufficient energy input altogether.

\subsubsection{Controlling to arbitrary non-uniform states}

\begin{figure*}[!ht]
\centering
\includegraphics[width=0.99\textwidth]{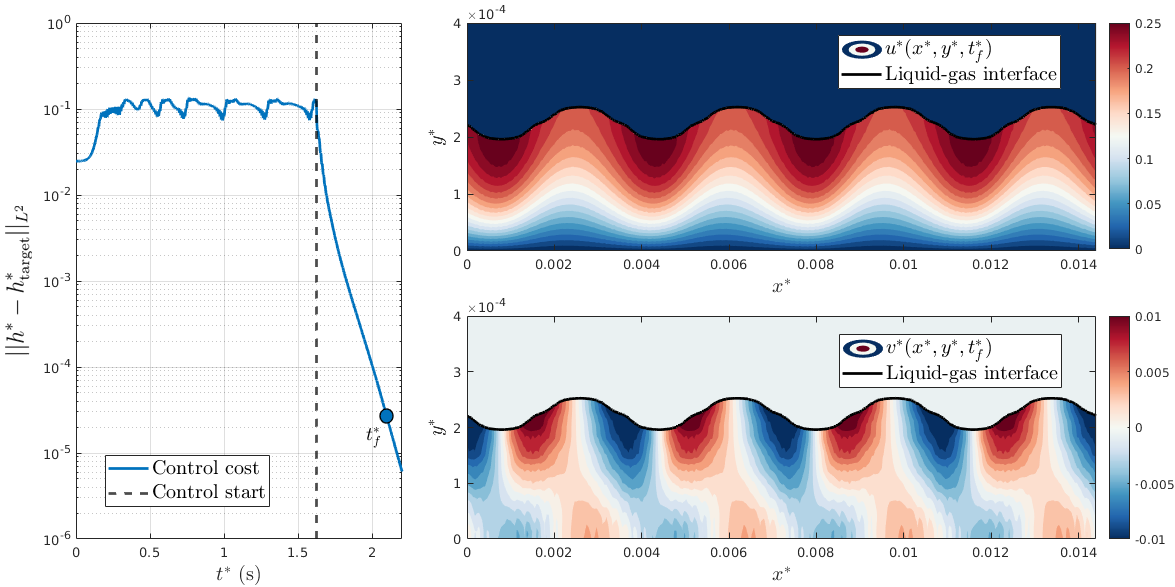}
\caption{Direct numerical simulation results for feedback control towards a non-uniform sinusoidal target solution $h^*_{\textrm{target}}(x^*)=\sin\left(\frac{\pi x^*}{8 h^*_0}\right)$
from the developed solution of a liquid film of undisturbed thickness $h_0^* = 225\ \mu$m falling down a plane with inclination angle $\theta = \pi/3$ using control strength $\alpha = 0.5$.
(Left) $L^2-$norm of the error between obtained and target interfacial shape. (Right) Horizontal velocity $u^*$ (top) and vertical velocity $v^*$ (bottom) at time $t^*_f \approx 2\ $s highlighted on the left hand side. The interfacial shape is highlighted in each subfigure, 
while the magnitude of each velocity component is restricted to the liquid.
Associated supplementary video material showcases convergence towards this target non-uniform state, as well as other imposed interfacial dynamics, for this particular set of physical parameters.}
\label{fig:nonuniform}
\end{figure*}

Up to this point, our results have have focused on the control towards a flat uniform state.
However, in many applications such as heat and mass transfer or directed assembly in micro-manufacturing, increasing interfacial area in 
a precise manner is a desirable feature. 
We have carried out extensive numerical computations to model a wide spectrum of scenarios, and focused on the control of both shapes and 
travelling wave velocities of the resulting 
interfacial profile. 
The control strategies implemented here were found to be successful as long as the target states do not deviate significantly from what
should be supported by the underlying physics. For example, surface tension will typically oppose interfacial shapes with small
wavelengths and large amplitudes, and we found that such cases
result in smoothed out shapes far from the desired target state. 
It is interesting to note that short-wave features that overcome such surface tension limitations can be robustly controlled using
externally applied electric fields, for example, with applications in micro- and nano-scale soft lithography \cite{schaffer2001electrohydrodynamic}.
However scenarios with primarily unimodal uncontrolled solutions can easily be switched to higher modes, as described for example by using a target 
sinusoidal interfacial shape with wavenumber $k=N \pi/L$. 
A DNS example using distributed controls with strength $\alpha = 0.5$ is shown in Figure~\ref{fig:nonuniform},
where we impose a target wave shape $h^*_{\textrm{target}}(x^*)=\sin\left(\frac{\pi x^*}{8 h^*_0}\right)$. 
Provided the actuation process is sufficiently strong, we found that prescribed solutions ranging from interfacial profiles effectively frozen in space 
(the shape no longer changes, but the bulk velocities are non-zero and the flow is moving down the plane subject to gravity), to an accelerated 
flow are all possible provided the energy input is deemed acceptable. 
Supplementary video material is provided for two such test scenarios, demonstrating the strength and flexibility of the proposed methodology.
In all cases the desired outcome was achieved with $\alpha = \mathcal{O}(1)$ 
and would in principle be a viable mechanism to specialise towards more specific applications.

The scenario summarised in Figure~\ref{fig:nonuniform} is highly nonlinear due to the relatively large interfacial perturbation amplitude. 
Nevertheless, following the initial transient, an exponential decay of the perturbation towards the target solution is observed (left panel - the
scales are logarithmic). 
Furthermore, the detailed analysis of the flow fields on the right hand side allows us to inspect features such as (i) a deceleration of the horizontal 
velocity near the peaks, compensated by an acceleration near the troughs, and (ii) propagation of information upstream from the
boundary to the interface as can be surmised from the vertical velocity plot in the lower right hand panel.
We have also found that, at sufficiently large values of $\alpha$ (depending on physical parameter values), 
local extrema in the interfacial profile align with the position of maximum blowing/suction at the
centre of the boundary strips. This type of patterning can be achieved at scale and is not restricted to a single 
type of surface profile. Microfluidic devices which require different features at different locations (or stages) 
could in principle take advantage of this flexibility.
Finally, we also note that while in the interface flattening scenarios the control will eventually be reduced close to (but never identically) zero,
in non-uniform scenarios it has to be maintained throughout in order to sustain non-trivial shapes.

Given the results above, it is natural to question whether we can control the system towards any designed steady state or saturated solution and what controls would be needed for that. The discussion in~\cite[Section V]{thompson2016stabilising} shows that we can reach any unstable steady state of our choice, with the caveat that these are model dependent and consequently the controls obtained cannot be readily ``translated" across the hierarchy of models. It is therefore suggested that it is more viable to control towards non-solutions $H(x)$. This can be achieved by adding a forcing term $S(x)$ that guarantees that $H$ is a (likely unstable) solution of the new forced equation, with the (full) control now being of the form
\[
F = -\alpha(h(x,t)-H(x)) + S(x).
\]
The \emph{constant in time} forcing $S(x)$ can be worked out by solving a linear PDE, in the case of a Benney equation, or a nonlinear PDE for the weighted residuals model -- 
these PDEs are obtained by replacing $H(x)$ in the respective model and obtaining a differential equation for $S$. Once $S$ is found, it is possible to show that 
the steady state $H$ is reached for \emph{large} $\alpha$ -- which in itself can already be an impediment to the controllability of the system. We note that the 
forcing $S(x)$ needs to be \emph{constantly} applied for the controls to work, even after the steady state $H$ is reached (otherwise $H$ is no longer a solution of 
the equation and the interface becomes unstable again). Therefore knowing $S$ can inform us of the feasibility of controls for a chosen $H$: if $S(x)$ is prohibitively 
large when compared to, e.g., the characteristic speed of the system, the controls are no longer ``imposed" but are rather ``enforced" and one could argue that they 
are not a realistic strategy.

\section{Discussion}
\label{s:conclusions}

Using the classical setup of a falling liquid film with imposed blowing/suction as active control mechanism,
we find that offline (or statically tailored) strategies provide a poor candidate for practical purposes due to the difficulty in synchronising 
exponentially decaying solutions with highly nonlinear systems. By contrast, feedback strategies (both distributed and point-actuated) provide 
a viable setup for convergence to desired states which can be simple/trivial (flat interfaces desirable in coating technologies) 
or spatially non-uniform (with increased interfacial area, of use in
heat and mass transfer applications, for example). 
The analytically informed high performance computing framework has been built on the basis of interrogating 
both stability and control theoretical results at the reduced-order model level. These included linear stability predictions on the minimum control strength $\alpha$ needed to stabilise a flat system, and on the best value for the spatial delay $\phi$ in the distributed case, complemented with the number of control actuators and their location for point actuated controls. The theoretical results provided {\it a priori} 
quantification on the behaviour of the controls which proved to be substantiated at the full model level. 
After confirming the validity of the model predictions, we could build on this knowledge and propagate similar strategies towards regimes outside 
the reach of the analytical (and numerical) methods. 

The particular interfacial flow setup used to demonstrate this interplay between techniques is a well-known and canonical one, 
however, more importantly, the approach has tremendous potential for generalisation towards complex systems in fluid mechanics 
and other fields. The toolkit has clear extensions in terms of classical control theoretical aspects, for example 
full feedback control, combining different observations to obtain a more efficient control rather than associating one observation to its own actuator, 
and using optimal control techniques to design controls which are robust to the hierarchy of models. Extensions in the application area include 
multi-physics scenarios such as electrohydrodynamics and soft lithography, heat and mass transfer, photo-excitation, and acoustics, to mention a few. 
In such applications the control design procedure is most often performed 
either very early (and thus with inaccurate propagation of information and insight) or very late (after expensive trial and error in early stages) in the process,
with the feedback control technology becoming of limited value, especially in view of further generalisation and applicability.

\section{Conclusion}
\label{s:summary}

Our study highlights a route towards the application of robust control strategies derived rigorously at the level of reduced-order 
mathematical models towards realistic ``in silico" experiments (presented in the form of direct numerical simulations). 
This allows, for the first time, the feasibility study of both offline and feedback approaches, alongside a systematic 
scrutiny of what system features are retained as the hierarchy in nonlinearity is traversed. 

Our proposed hybrid control methodology harnesses multiple modelling capabilities, unblocking typical individual approach bottlenecks. This has significant potential in enabling our wider research community to efficiently investigate not only new regimes (as exemplified here through a classical and highly nonlinear problem in interfacial fluid mechanics), but also new physical mechanisms and applications which have previously proven inaccessible. In particular, it:
\begin{enumerate}
 \item is unique in its multi-faceted approach towards efficient modelling, with both analytical and computational components acting in tandem,
 \item safeguards against the overly simplistic use of offline control methods,
 \item offers a paradigm shift in terms of control information transition between modelling approaches,
 \item informs viable validated control strategies within the range of applicability of analytical methods, and
 \item provides access to wide regions in parameter space, and is a natural departure point for feedback control applications even 
in cases beyond the validity of the asymptotic models, with capabilities of informing a powerful integrated computational platform and ensuring a step change in 
transitioning towards real-life systems.
\end{enumerate}

The hierarchical control design approach is anticipated to considerably reduce expensive experimental and manufacturing pipelines 
and is well suited towards specialisation within a wide variety of physical, engineering and cross-disciplinary contexts. 

\section*{Declarations}
\textbf{Funding} R.C. gratefully acknowledges the resources and continued support of the Imperial College London Research Computing Service.
He also underlines the role of the University of Oxford Mathematical Institute funding in the form of the Hooke Research Fellowship,
as well as resources from the UK Fluids Network (EPSRC Grant EP/ N032861/1), 
which have enabled significant cross-institutional interactions. S.N.G. is supported by the Leverhulme Trust through the Early Career Fellowship ECF-2018-536. 
D.T.P. was partly supported by EPSRC Grant EP/L020564/1.

\medskip
\noindent
\textbf{Conflicts of interest} The authors declare that they have no conflict of interest.

\medskip
\noindent
\textbf{Author contributions}  All authors contributed to the study conception and design. R.C. and S.N.G. performed the research, with R.C. primarily developing direct numerical simulation capabilities, S.N.G. developing reduced-order modelling numerical solvers and associated tools, and both co-authors contributing equally in terms of unified software implementation for comparison. All authors were involved in interpretation discussions, as well as writing, reviewing and editing.

\appendix
\section{Parameter values}
\label{A:parameters}
In all our numerical calculations we considered parameters pertaining to a water-air system which is easily realisable experimentally 
(see e.g. \cite{denner2018solitary,kalliadasis2011falling}). 
We stress, however, that the approaches discussed here are applicable to liquid-gas systems in general, 
as long as the assumptions behind the modelling framework hold. 
The inclination angle is fixed at $\theta = \frac{\pi}{3}$ and the fluid thickness is varied from $h_0^* = 125\ \mu$m to $h_0^* = 225\ \mu$m,
recalling that the $(\cdot)^*$ notation is used to denote dimensional quantities.
The periodic domain is constructed with an aspect ratio of $64:1$, such that $L^* = 64h_0^*$, which is in accordance with the long-wave assumption. 
These settings translate to \emph{moderate} Reynolds number flows where the uncontrolled dynamics exhibits a range of behaviours, including nonlinear travelling waves, trains of solitary wave pulses, three-dimensional dynamics and spatiotemporal chaos.
The thinnest film $h_0^* = 125\ \mu$m is
just beyond the threshold for linear stability and hence resulting 
in the formation of mildly nonlinear travelling waves, 
while the upper limit $h_0^* = 225\ \mu$m pertains to sufficiently thick films in more inertially-dominated flow regimes so that 
the reduced-order models start losing their validity and computations become restrictive due to numerical stiffness.

Relevant parameters for the liquid are its density $\rho^*_l = 998\ kg/m^3$ and dynamic viscosity $\mu^*_l = 8.967\times 10^{-4}\ kg/ms$, 
while the equivalent gas parameters are $\rho^*_g = 1.17\ kg/m^3$ and $\mu^*_g = 1.836\times 10^{-5}\ kg/ms$.
The water-air surface tension coefficient is taken to be $\gamma^* = 0.072\ kg/s^2$, while the acceleration of gravity is given by $g^* = 9.807\ m/s^2$. 
For the ranges of interface heights we consider in our numerical experiments, we obtain the reference velocities and dimensionless groupings 
presented in Table~\ref{tab:values} below, with the liquid values considered as reference (see details in subsection~\ref{ss:ROM}).
\begin{table}[h]
	\begin{center}
	\begin{tabular}{cccc}
		$h_0^* \ (\mu m)$ & $U_s^* \ (m/s)$ & $Re$ & $Ca$ \\
		\hline
		125 & $0.0738$ & $10.2739$ & $0.0009$\\
		150 & $0.1063$ & $17.7532$ & $0.0013$ \\
		175 & $0.1447$ & $28.1915$ & $0.0018$ \\
		200 & $0.1891$ & $42.0817$ & $0.0024$ \\
		225 & $0.2393$ & $59.9171$ & $0.0030$\\
	\end{tabular}
	\end{center}
\caption{Parameter values used in our numerical experiments: interface height and corresponding surface speed, Reynolds number $Re$ and Capillary number $Ca$.}
\label{tab:values}
\end{table}

\section{Numerical solution of the reduced-order models}
\label{B1:PDEsol}
The solutions to the reduced-order models presented in Figure~\ref{fig:validation} were obtained as follows. The Kuramoto-Sivashinsky equation was solved in an appropriately rescaled periodic domain $x \in [0,2\pi]$ using a pseudo-spectral method for the spatial discretisation based on a Fourier series decomposition of the solution. 
To time-step, we used second order backward differentiation formulae (see~\cite{akrivis2011linearly}) 
with a sufficiently small time step. The weighted residuals model was solved using a pseudo-spectral spatial discretisation and a variable-step, variable-order time-stepping scheme based on the numerical differentiation formulas (NDFs) of orders 1 to 5, provided by \textsc{Matlab}'s inbuilt function \emph{ode15s}.

\section{Direct numerical simulations and validation}
\label{B:DNSdetails}

We have developed the code for the direct numerical simulations conducted in this investigation as part of the open-source solver 
$\mathpzc{Gerris}$ \cite{popinet1,popinet2}. The main implementation, which can be found at \texttt{http://gfs.sourceforge.net}, 
is ideally sui- ted for interfacial flows in both fundamental and applied contexts and has seen adoption and growth over the past two decades.

Here we briefly sketch the approach for our particular problem and refer the reader to~\cite{popinet2} for further details.
The equations of motion are 
\begin{align}\label{eq:GerrisEquations}
 \rho(\partial_t \textbf{u}_f+\textbf{u}_f\cdot \nabla \textbf{u}_f) &= -\nabla p + \nabla \cdot(2 \mu\boldsymbol{\mathcal{D}}) + \sigma \kappa \delta_s \textbf{n}+\textbf{F}_e, \nonumber \\ 
 \partial_t \rho + \nabla \cdot (\rho \textbf{u}_f) &= 0, \\ 
 \nabla \cdot \textbf{u}_f &= 0, \nonumber
\end{align}
where $\boldsymbol{\mathcal{D}}$ is the rate of strain tensor
$\mathcal{D}_{ij} =(1/2) (\partial u_i/\partial x_j +\partial u_j/\partial x_i)$.
We stress that the full Navier-Stokes equations are solved in both the liquid and the gas, with the decoration $(\cdot)_f$ being a placed holder 
for both gas-related - $(\cdot)_g$ - or liquid-related - $(\cdot)_l$ - quantities.
In the interfacial region between the two, discrete counterparts 
of the normal and tangential stress balances, the kinematic condition, as well as continuity of velocities are imposed. All interfacial forces are transferred to the
momentum equations in what is commonly called the ``one-fluid" formulation.
The physical properties describing each fluid (density, viscosity, permittivity etc.) are included by singular 
distributions and the same set of equations~\eqref{eq:GerrisEquations} accounts for the entire domain.
The Dirac distribution $\delta_s$ isolates the surface tension effects to the interface alone, and
any external forces such as gravity are included via the $\textbf{F}_e$ term. 

The relevant physical properties of the fluids (such as density and viscosity) are represented in terms
of a volume fraction $c(\textbf{x},t)$, where $c$ is a generic colour function which takes the
value $0$ in one fluid (e.g. the gas) and $1$ in the other (e.g. the liquid, our reference fluid). More specifically we write
\begin{equation}
 \rho(c) \equiv c \rho_l + (1-c) \rho_g,\quad  \mu(c) \equiv c \mu_l + (1-c) \mu_g, \label{eq:propertiesGerris}
\end{equation}
Under this treatment, a density equation (the other properties are treated in a similar manner) of the general form
\begin{equation}
 \rho_t + \nabla \cdot (\rho \textbf{u}_f) = 0
\end{equation}
becomes
\begin{equation}
 c_t+\nabla \cdot (c \textbf{u}_f) = 0,
\end{equation}
which is solved for $c$ and the results substituted into~\eqref{eq:propertiesGerris}. 
The value of $c$ is interpolated across the interface by introducing a small transition
layer in its vicinity to smooth the variation of quantities from one region to the other, thus relaxing an otherwise singular transition.

The discretisation schemes are $2^{\textrm{nd}}$-order adaptive in both time and space, with a series of criteria ranging from refinement settings to capillarity- 
and inertia-based constraints dictating the required timestep to advance the solution within a tolerated error bound. The critical numerical step 
in the process is a multi-level Poisson solver for pressure (following projection steps). Efficient solutions thereof are facilitated by 
excellent spatial adaptivity which can span $3$ orders of magnitude in cell sizes, thus reducing the number of degrees of freedom considerably and allocating 
them where needed. In our particular case, we employed changes in the velocity fields and vorticity, as well as interfacial location as refinement criteria, 
leading to systems with $\mathcal{O}(10^4)$ cells. Furthermore, the structured mesh based on an underlying quadtree (octree in 3D) setup results 
in strong parallelisation properties. Most numerical experiments have been run on $4-16$ CPUs on local high performance computing facilities 
at Imperial College London and required roughly $100-1000$ CPU hours depending on the maximum allowed resolution level.

From each individual simulation we extract a rich dataset consisting of interfacial heights, virtual measurement station output, 
as well as complete flow snapshots encapsulating flow quantities such as velocity fields and pressure for the entire 
domain or in a desired localised area. 
Each multi-level and multi-frequency output procedure results in 
thousands of files with dedicated post-processing scripts and characterised by at least $1\ $Gb in size. While not prohibitive 
at an individual level, full parameter studies can quickly become very demanding even for large scale computing clusters. This is the primary reason why model-informed 
computations are not only desirable, but very much a necessity in order to make efficient use of resources.
The ultimate advantages of deploying this advanced computational machinery lies in the ability to investigate the full physics (e.g. high-order nonlinearities) 
which were removed by the modelling assumptions in order to make analytical progress. 
Furthermore, this grants us the ability to move deeply into regions of parameter space that are not hindered by the restriction of disparity in lengthscales
required by the models, or result in difficult scenarios such as multi-valued interfaces and topological changes.
With the aid of the model, the system stability properties, and the control behaviour, 
such lines of investigation become genuinely tractable in terms of rigorous study as opposed to single case-study type parameter examinations.

We have conducted an extensive validation exercise based on two major strategies. Firstly, we have employed 
numerical criteria such as ensuring mass conservation and mesh-independence of the results (to within an accuracy of $\mathcal{O}(10^{-6})$ in the constructed norms 
pertaining to $\mathcal{O}(1)$ changes in the flow properties), 
which allowed initial resolution and refinement criteria tailoring.
Secondly, the reason for choosing this particular application lies in the fact that both model solutions (re-visit subsection~\ref{ss:ROM}) and previous
numerical and experimental solutions, e.g. \cite{denner2018solitary}, are available for comparison. The former were employed as stringent target cases for relatively 
small thicknesses were the models are known to be valid, while the latter resources allowed additional testing in more challenging and often inertially-dominated 
regimes outside the reach of the models.

The summary in Figure~\ref{fig:validation} pertains to a scenario which constitutes the most difficult test case for which modelling efforts still converge to a solution. Similar studies have been performed for smaller film thicknesses. 
We noticed that the saturated nonlinear wave shapes compared very well in all such cases and that the evolution itself (based on tracking the interfacial 
height at specific measurement stations) matches surprisingly well even at moderate Reynolds numbers (of up to $Re \approx 30.0$).
There are however small but non-negligible differences in both the amplitude of the final interfacial shape and the 
periodicity of the resulting dynamics. Thus, small adjustment factors accounting for errors within $5\%$ were necessary in order to synchronise 
an exact start to any control procedure for offline controls. Feedback controls were unaffected by such features 
and subsection~\ref{ss:feedback} outlines the excellent agreement between model and full DNS predictions through the discussions 
around Figures~\ref{fig:fullsurface}-\ref{fig:pointactuated}.

\bibliographystyle{spmpsci}      
\bibliography{FallingFilmBib}   

\end{document}